\documentclass[preprint]{vgtc}                          

\graphicspath{{figures/}{pictures/}{images/}{./}} 

\usepackage{times}                     

\usepackage{tabu}                      
\usepackage{booktabs}                  
\usepackage{lipsum}                    
\usepackage{mwe}                       
\usepackage{tikz}

\usepackage{mathptmx}                  

\usepackage{subfig}
\usepackage{csquotes}
\usepackage[flushleft]{threeparttable}
\usepackage{float}
\usepackage{comment}

\usepackage{balance}

\onlineid{0}

\vgtccategory{Research}

\title{Working with Mixed Reality in Public: 
Effects of Virtual Display Layouts on Productivity, Feeling of Safety, and Social Acceptability}

\author{Janne Kaeder\thanks{e-mail: janne.kaeder@gmx.de}\\ %
    \parbox{2.0in}{\scriptsize \centering Technische Universität Berlin, Germany}
\and Maurizio Vergari\thanks{e-mail: maurizio.vergari@tu-berlin.de}\\ %
     \parbox{2.0in}{\scriptsize \centering Technische Universität Berlin, Germany}
\and Verena Biener\thanks{e-mail: verena.biener@visus.uni-stuttgart.de}\\ %
     \parbox{2.0in}{\scriptsize \centering University of Stuttgart, Germany}
\and Tanja Kojić\thanks{e-mail: tanja.kojic@tu-berlin.de}\\ %
     \parbox{2.9in}{\scriptsize \centering Technische Universität Berlin, Germany}
\and Jens Grubert\thanks{e-mail: jens.grubert@hs-coburg.de}\\ %
    \parbox{2.9in}{\scriptsize \centering Coburg University of Applied Sciences and Arts, Germany}
\and Sebastian Möller\thanks{e-mail: sebastian.moeller@tu-berlin.de}\\ %
    \parbox{2.9in}{\scriptsize \centering Technische Universität Berlin \& DFKI, Germany}
\and Jan-Niklas Voigt Antons\thanks{e-mail: jan-niklas.voigt-antons@hshl.de}\\ %
    \parbox{2.9in}{\scriptsize \centering Hochschule Hamm-Lippstadt, Germany}}

\teaser{
\vspace{-3.7mm}
\centering
\includegraphics[width=0.246\textwidth, height = 2.8cm]{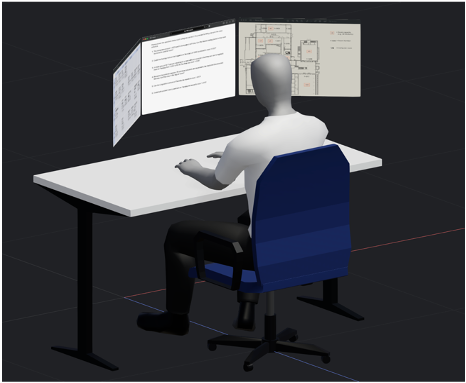}
\includegraphics[width=0.246\textwidth, height = 2.8cm]{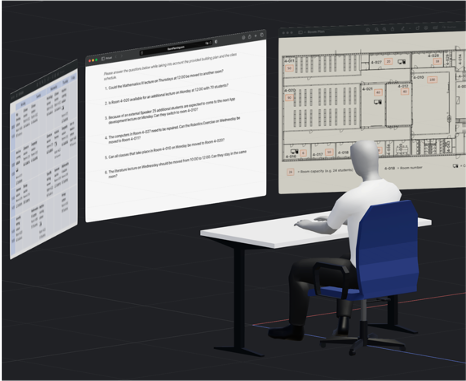}
\includegraphics[width=0.246\textwidth, height = 2.8cm]{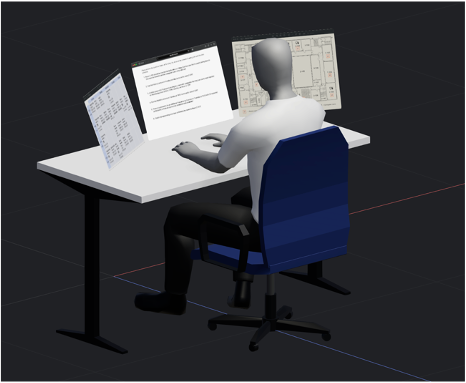}
\includegraphics[width=0.246\textwidth, height = 2.8cm]{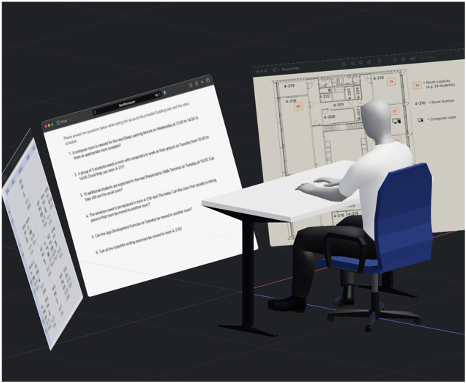}
\caption{The four different layouts, from left to right: eye level-close (EL/C), eye level-far (EL/F), below eye level-close (BEL/C), below eye level-far (BEL/F)}
\label{fig: teaser}
}

\abstract{
Nowadays, Mixed Reality (MR) headsets are a game-changer for knowledge work. Unlike stationary monitors, MR headsets allow users to work with large virtual displays anywhere they wear the headset, whether in a professional office, a public setting like a cafe, or a quiet space like a library. This study compares four different layouts (eye level-close, eye level-far, below eye level-close, below eye level-far) of virtual displays regarding feelings of safety, perceived productivity, and social acceptability when working with MR in public. We test which layout is most preferred by users and seek to understand which factors affect users' layout preferences. The aim is to derive useful insights for designing better MR layouts. A field study in a public library was conducted using a within-subject design. While the participants interact with a layout, they are asked to work on a planning task. The results from a repeated measure ANOVA show a statistically significant effect on productivity but not on safety and social acceptability. Additionally, we report preferences expressed by the users regarding the layouts and using MR in public.
} 


\begin{document}
\raggedbottom


\maketitle

\section{Introduction}

In the past years, technological advancements have been made in Extended Reality (XR). A notable trend is that Virtual Reality (VR) headsets have been increasingly equipped with passthrough capabilities, which enable users also to experience video see-through Mixed Reality (MR). Through cameras and sensors in the device, video passthrough allows users to view their physical environment while wearing the headset. As MR allows for situational awareness, these headsets become more suitable for public use. Another major benefit of MR is that it allows users to position screens of almost arbitrary size in any environment. This feature is especially relevant for knowledge work since many people are used to working with external monitors at home or in the office. In contrast to stationary monitors, MR allows users to work with an array of large virtual displays wherever they wear the headset, for example, in public transport, at a café or the library. Therefore, we might start to see people using XR headsets in public settings. For example, when Apple announced its \enquote{Vision Pro} headset, its promotional video showcased how it could be used on a plane, following earlier visions from academia (e.g., \cite{grubert2018office}). As MR technology transitions from private to public settings, it's crucial to consider additional factors such as distractions, social acceptability, and safety, all of which significantly impact the user experience (UX). While extensive research has been conducted to enhance the UX of VR and MR, it remains uncertain whether the design recommendations derived from these studies are applicable in public settings. This underscores the necessity for more research to delve into how users envision using MR in public. Such studies are instrumental in identifying user preferences for virtual screen layouts when utilizing XR headsets in public spaces.

\section{Related Work}
\label{sec: related_work}

\subsection{Social Acceptability, Safety and Productivity of XR in public}

Due to the rising popularity of XR headsets in the past years, research dedicated to using XR in public spaces has gained increasing attention. When immersive technologies are transferred from private to public settings, additional factors like distractions, social acceptability, and safety influence the user experience and need to be considered.

Social acceptability is crucial for the adoption of new technologies like immersive headsets. It can be explored from both the bystanders' and the users' perspectives. Alallah et al. \cite{alallah2018performer} studied the social acceptability of input modalities for head-mounted displays, finding that the visibility of a gesture influences both performer and spectators’ acceptance. A survey by Schwind et al. \cite{schwind2018virtual} that focuses on the spectator's perspective found that the social acceptability of wearing an XR headset depends on the situation and the environment. In settings like in the bedroom, the metro, or on a train, the social acceptability is significantly higher than in settings where it is expected to engage in social interactions like in the living room or at a public café. Bajorunaite et al. \cite{bajorunaite2021virtual} explored the acceptability of using VR in different modes of public transport, like on a bus, a train, or an airplane. They found that people are more interested in using VR on longer journeys. They identified that the main barriers to VR adoption in public were concerns about loss of awareness and being stared at or judged by other passengers not wearing a headset. A recent follow-up study by the same researchers \cite{bajorunaite2023reality} investigated how the acceptance of VR in public transport could be increased by enriching the immersive experience with visual cues about people and objects in their physical surroundings, which they call \enquote{Reality Anchors.} They conducted a lab study that simulated a public transport journey in VR and found that the visibility of Reality Anchors within an immersive experience made users judge the experience as more socially acceptable. Additionally, users felt safer when they saw cues from reality in VR but also tended to be more distracted. The visibility of other people and personal belongings was especially important for users.

Previous research found that safety is a major concern when using VR in public \cite{bajorunaite2021virtual,eghbali2019social}. A recent study by Biener et al. \cite{biener2024} supports the finding that users feel safer when they can see their surroundings. In this study, participants were seated in a public space at a university, and three different setups for knowledge work were compared: working on a laptop, in MR, and in VR. The results showed that participants felt the safest when working on the laptop and significantly safer in MR than VR. 

Further research about productivity includes a study by Lee et al. \cite{lee2022partitioning}, which explored how AR could reduce distractions in open-plan workspaces by projecting virtual separators around a user's desk. They found that the virtual partitions were effective in reducing visual distractions. Additionally, they compared different designs of virtual partitions. The results showed that most participants preferred using semi-transparent partitions over opaque ones despite being less effective in blocking visual disturbances.

\subsection{Layouts of virtual displays}

Two studies by researchers from the University of Glasgow \cite{medeiros2022shielding, ng2021passenger} investigated preferences for layouts of virtual screens in MR in different shared transport settings. For that purpose, VR was used to simulate different public transport settings, like planes, trains, or subways. They found that different environments affect user's preferences for layouts. For example, in a narrow space, like on a plane, people prefer a vertical arrangement of screens because they do not want to invade the personal space of the people sitting next to them, while in other settings, a horizontal layout is preferred. Research by Cheng et al. \cite{cheng2021semanticadapt}, further supports the finding that users prefer MR layouts that do not invade other people's personal space. When users are given complete freedom to place virtual displays according to their preferences, it has been observed that people try to avoid the virtual displays collide with objects in their physical surroundings. Instead, people use physical surfaces, like walls and tables, to align the displays. 
Additionally, Medeiros et al. \cite{medeiros2022shielding}, identified the trend of people avoiding placing displays over other passengers or their belongings. Participants' reasoning indicates that this is due to situational awareness and social norms. Users want to be aware of what is happening in their surroundings and do not want to give the impression that they are staring at other people or their belongings. The qualitative results show that acting by social norms is a high priority for users and, in many cases, is considered more important than safety or physical comfort. However, some participants prioritized ergonomic comfort and placed the displays before other passengers. The fact that participants did not interact with the content of the virtual displays for a prolonged period limits this study. The authors acknowledge that ergonomic factors could influence the layout preferences when interacting with the displays for a longer period. Furthermore, since the study was conducted within a VR simulation of public transport settings with other passengers depicted as avatars, it remains to be seen how the results might vary in a real-world setting with real observers.

In order to help users create satisfying MR layouts across different contexts, it has been suggested to adapt layouts based on the current environment automatically \cite{cheng2021semanticadapt, cheng2023interactionadapt,lindlbauer2022future, medeiros2022shielding,ng2021passenger}. For instance, an algorithm that senses and adapts to the user's context could automatically rearrange the virtual displays when a user boards a train or a person is sitting down across from them.

Next to the body of research specifically focused on XR in public spaces, most studies exploring layout preferences are centered around XR use in private settings. For example, Kojic et al. found that users prefer flat canvases over curved ones \cite{kojic2022assessing}. Regarding the distances of virtual screens, a larger distance was preferred by participants and allowed them to read text faster compared to displays placed at a smaller distance. It remains to be seen how applicable these results are in public settings, where displays placed at further distances might invade bystander's personal space.

\subsection{Objectives}

This study explores layout preferences for working in MR in a real-world setting (a public library). It compares different layouts of virtual displays concerning feelings of safety, perceived productivity, and social acceptability.

For this purpose, four different layouts are tested (see Fig. \ref{fig: teaser}). The different layouts vary in how much they cover observers in the room. Layouts EL/C and EL/F are positioned at eye level and will cover people sitting across from the user. Layouts BEL/C and BEL/F are positioned lower so the user can still see other people sitting across from them. Furthermore, the layouts vary regarding whether they invade bystanders' personal space. The displays of layouts EL/C and BEL/C are positioned close to the user and within the area above the user's desk. In contrast, layouts EL/F and BEL/F are positioned further away and extend to the desk of the person sitting across from them. In the related studies presented on layouts in MR, participants frequently mentioned the factors of social acceptability and safety when expressing layout preferences. Additionally, productivity is investigated as the presented layouts are supposed to be suitable for knowledge work. The first objective of this study is to compare the different layouts of virtual displays regarding feeling of safety, perceived productivity, and social acceptability. Social acceptability is only examined from the user's perspective. The second objective is to identify which layout is most preferred by users in this specific public setting and understand which factors affect user's layout preferences. The aim is to derive insights that contribute to informing the design of tools that help users create satisfying layouts.

\section{Methods}

\subsection{Research Questions \& Hypotheses}

Based on the related work and objectives (see Sec.\ref{sec: related_work}), the following hypotheses (H) and research questions (RQ) are formulated for public workspaces at a library:

\begin{description}
    \vspace{-1 mm}
    \item[H1:] Layouts with displays below eye level (BEL) make users feel safer than layouts with displays on eye level (EL).
    \vspace{-2mm}
    \item[H2:] Layouts with displays on eye level (EL) make users feel more productive than layouts with displays below eye level (BEL).
    \vspace{-2mm}
    \item[H3:] Layouts with displays below eye level (BEL) are perceived as more socially acceptable by users than layouts with displays on eye level (EL).
\end{description}

\begin{description}
    \vspace{-4mm}
    \item[RQ1:] Which of the four layouts is the most preferred by participants?
    \vspace{-2mm}
    \item[RQ2:] What are user's reasons for their layout preferences?
\end{description}

\subsection{Experimental Design}

The study follows a single-factor experimental design, with the layout of the virtual displays as the independent variable and the feeling of safety, perceived productivity, and perceived social acceptability as the dependent variables. The layout variable has four levels (EL/C, EL/F, BEL/C, and BEL/F), and each participant experiences all four layout conditions. It must be noted that the layouts do not differ only in height and distance, but other factors like angle of viewing and size were tweaked (see Sec. \ref{sec: test setup}). For this reason, the layout is considered to be a single independent variable with four different levels. A within-subject design was chosen to better control for individual differences. For variables like social acceptability and feeling of safety, participants are likely to have individual attitudes regardless of the layouts. Having participants rate their experience for all layouts reduces the influence of these individual differences on the results. The order of the layouts was counterbalanced to avoid sequence effects. A quantitative research approach is used to test the hypotheses. The dependent variables are measured through questionnaires administered after each layout condition. The questionnaires are taken from other studies and were slightly adapted to fit the context of this study. Since the questionnaires have already been proven successful in measuring feelings of safety, productivity, social acceptability and comfort in other studies \cite{pavanatto2021we, bajorunaite2023reality, eghbali2019social, vergari2021influence, mcgill2020expanding}, they are expected to measure the dependent variables accurately and reliably. Additionally, qualitative data is gathered through open-ended questions in the final questionnaire. This data can be used to explain the quantitative results and better understand the participants' subjective experiences. 

When investigating the use of MR in public, the environment in which the study takes place is an essential part of the study design. In order to explore the experience of using MR in a real social setting and achieve a high ecological validity, the study is designed as a field study. A public library was chosen for the study because it represents an ideal example of a public working space. This is an interesting environment to explore, as working in public workspaces has become very popular in different forms \cite{krauss2024third}. Such environments are suitable for using XR headsets, which enable working with multiple large screens, which might otherwise be impractical. The tables were arranged in rows, so people sat across each other.

Conducting the study in a real public setting also means that the environment is difficult to control, which could influence the measured variables. Generally, a trade-off between the internal and external validity of the study had to be made. Since a lack of research that investigates the use of MR in real public settings has been identified, some control over the environment is sacrificed in favor of higher external validity. Nevertheless, measures are taken to improve the internal validity. For example, the study is conducted only on weekdays between 10:30 and 6 pm when other people usually occupy the study space. Since the library tends to get empty during lunchtime, it is avoided to conduct the study between 12 and 2 pm. Additionally, it is ensured that there is always a person sitting across from the participant because this is expected to impact the experience with different layouts.

An a priori power analysis was conducted using the software G*power \cite{faul2007g} to determine the required sample size to test the study's hypotheses using repeated measures ANOVA. When assuming a correlation among repeated measures of 0.5, the result of the power analysis showed that, for detecting a medium effect (f = 0.3) with an $\alpha$ error probability of 0.05, a sample size of N = 17 is needed to achieve a power of 0.80.

\subsection{The Room Planning Task}

While the participants interact with a layout, they are asked to work on a room planning task. The room planning task was designed by Biener et al. to measure productivity \cite{biener2024}. The same task was adapted to fit the context of this study. 

The room planning task puts the participant in the role of a university employee, who is responsible for coordinating which courses should take place in which rooms of a university building. The participant is presented with two documents: a class schedule and a room plan. Based on the provided information, they must answer room planning-related questions, for example, if a specific lecture can be moved to a different room. The task was simplified in terms of the amount of information presented in the room plans and class schedules as well as the complexity of the questions, which were rephrased to \enquote{yes} or \enquote{no} questions.

The participant is presented with a room plan document, a course schedule, and six room planning questions for each layout condition. The course schedule contains the name of each course, the day and time of each course, the number of attending students, and whether a computer is required for the course, as indicated by a computer icon. The room plan contains for each room: the room number, the room capacity (number of students it can fit), and a computer icon if the room is a computer room. To determine if a room change is possible, the participant must consider whether the mentioned room is already occupied on the specific day and time, whether the room is big enough to fit all students, and, if the course requires computers, consider the computer availability in the room.

For each layout condition, participants are expected to take between three to five minutes to answer the six questions, depending on their speed. The room planning questions are designed to be equally difficult to compare the perceived productivity across the different layout conditions. To reduce learning effects across the four conditions, not only do the room planning questions differ for each condition, but also the room plan and class schedule. This task was chosen because it is a knowledge work task that requires concentration and is expected to be challenging enough to measure perceived productivity. Furthermore, the task requires participants to combine information from all three virtual displays of a layout, which ensures that participants interact with all parts of the layout.

\subsection{Test Setup}

\label{sec: test setup}

The developed application\footnote{github.com/jannekaeder/MR-Prototypes-of-Virtual-Display-Layouts} was running on a Meta Quest 3 headset. The Meta Quest 3 provides all necessary MR functionalities. Oculus Touch controllers were used as the input modality for the MR application. The MR application was developed using Unity, Bezi\footnote{bezi.com}, and Figma. Bezi is a platform that allows designers to prototype 3D applications quickly. Furthermore, Bezi offers integrations for the popular design tool Figma and Unity. Therefore, frames designed in Figma can be imported into Bezi, and Bezi prototypes can be exported to Unity.

\begin{figure*}[h!]
\centering
\includegraphics[width=0.33\textwidth]{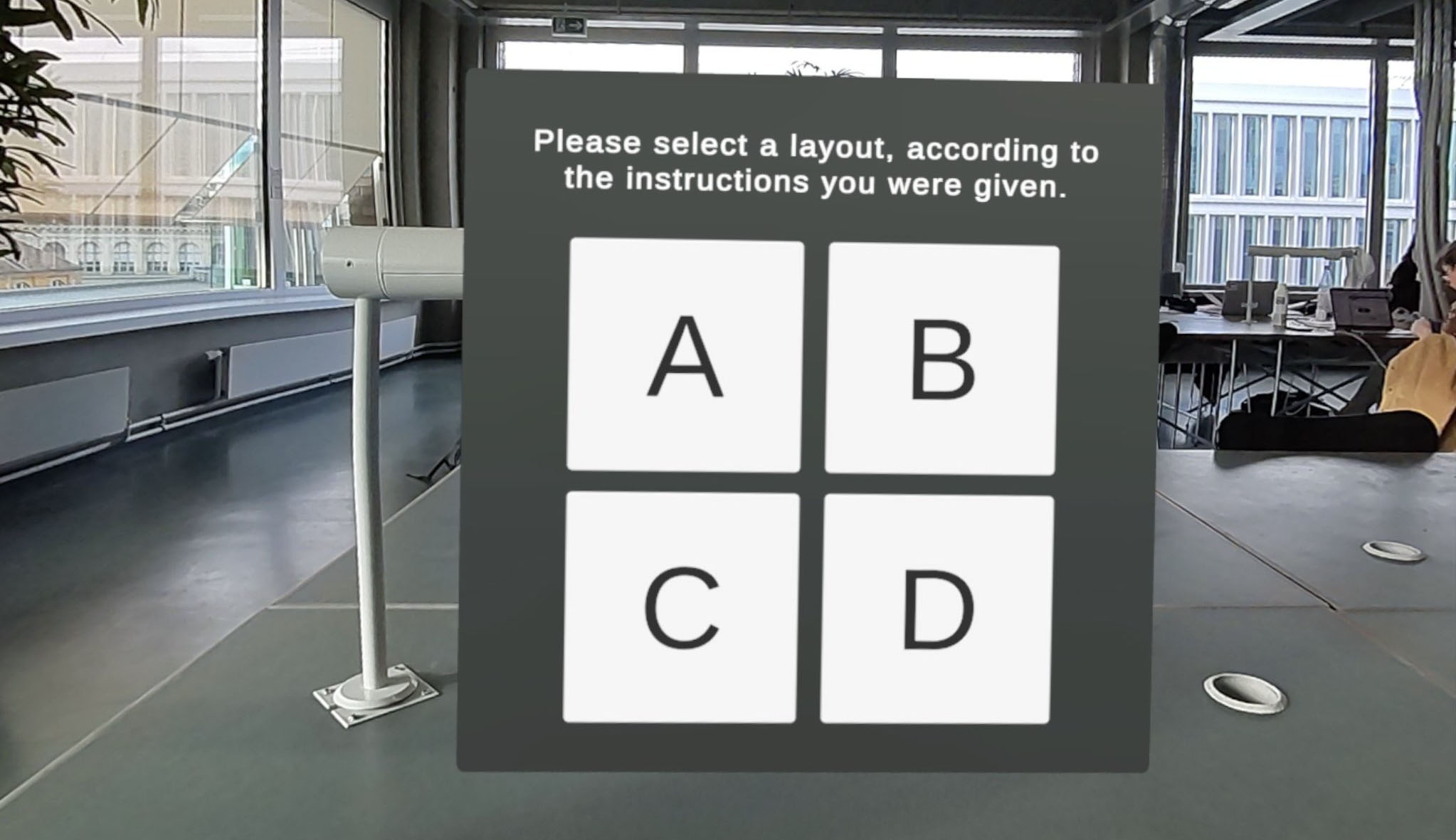}
\includegraphics[width=0.33\textwidth]{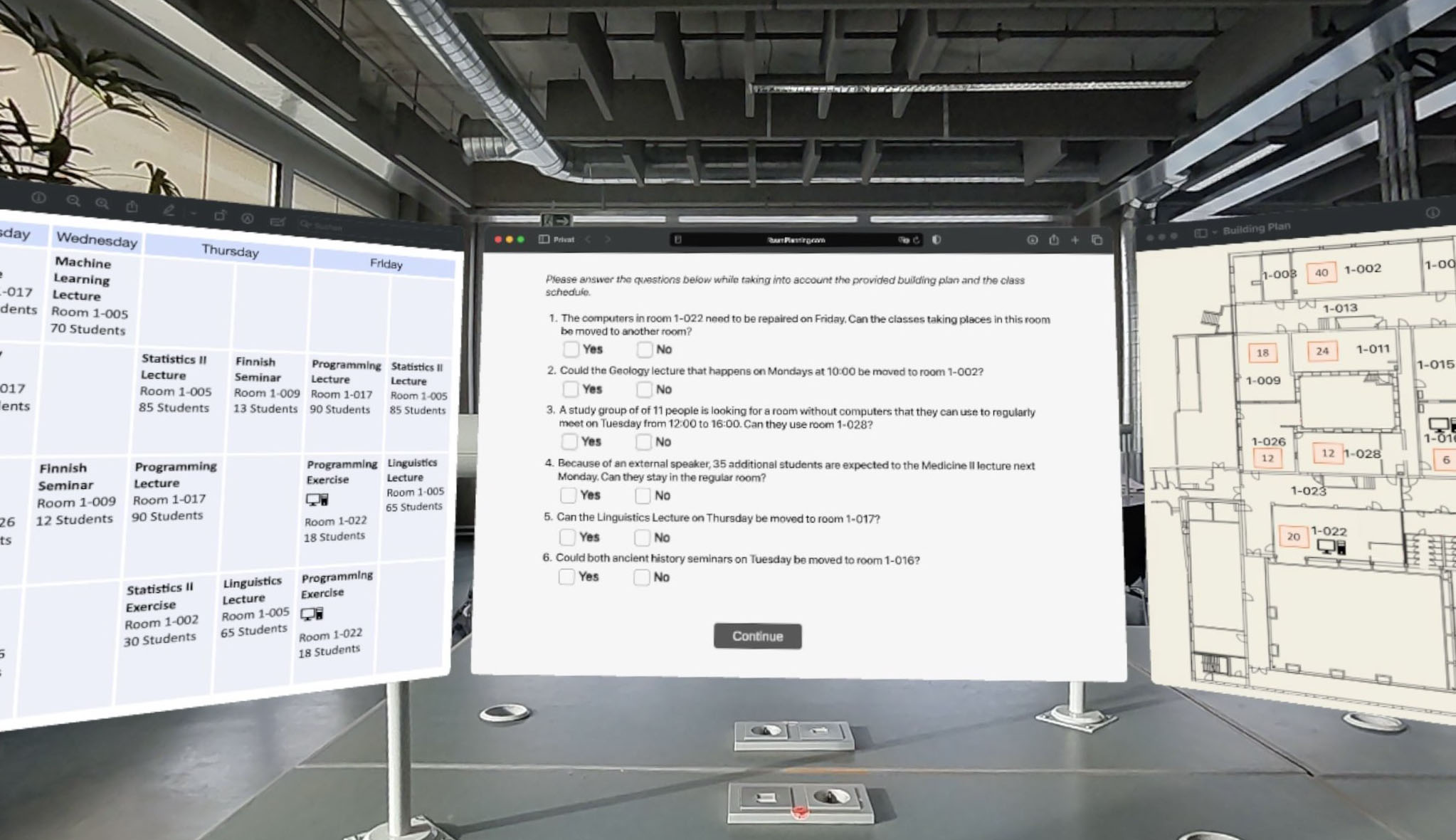}
\includegraphics[width=0.33\textwidth]{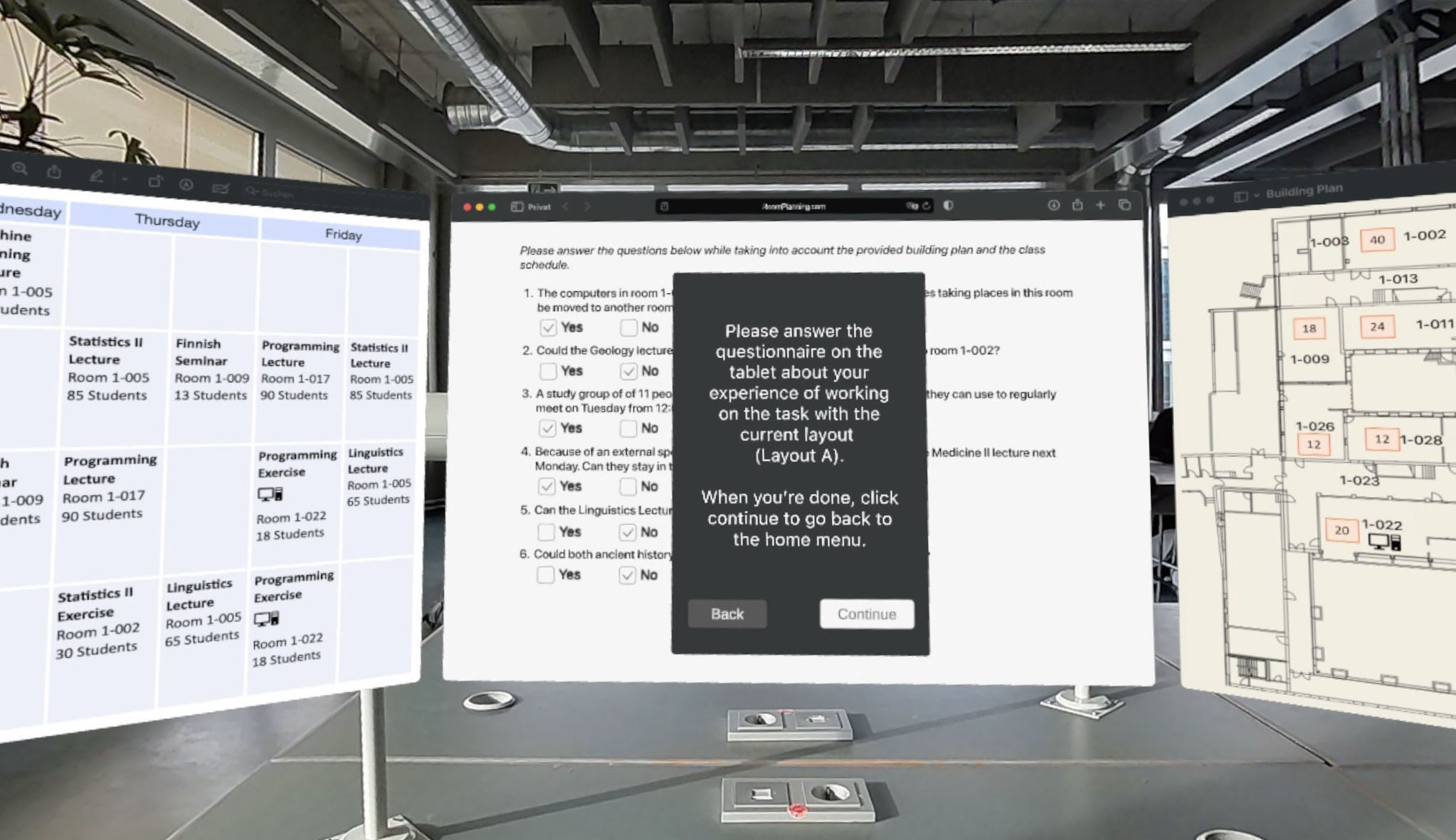}

\caption{Prototype flow, from left to right: home menu, room planning task, pop-up window}
\label{fig: prototype}
\vspace{-0.3cm}
\end{figure*}

When creating the four different layouts, important design decisions about the height, angles, distances, and sizes of the virtual displays had to be made. When arranging multiple virtual displays next to each other, it is common practice to rotate them to facilitate an easier view of all the displays \cite{mcgill2020expanding}. Therefore, the side displays in all layouts were rotated by 30 degrees so that all displays are oriented towards the user. Furthermore, it is common for displays placed below eye level to be tilted towards the back. For example, the operating system of the Meta Quest 3 always arranges such displays at an angle of 70 degrees. Therefore, the same was done for the displays in the layouts below eye level. Regarding the displays' height, guidelines by Meta and Microsoft recommended not placing the center of the displays exactly on eye level, but rather slightly below, to allow for a more comfortable head position. This recommendation was followed, and the center of the virtual displays were placed slightly below eye level in the layouts on \enquote{eye level} (layouts EL/C and EL/F)  while also ensuring that the lower edge of the displays does not collide with the desk at which the user is sitting. The height of the displays in the layouts below eye level was set so that the face of the person sitting across from the user was visible above the upper edge of the central display. Regarding the distance of virtual objects, guidelines by Meta recommend placing content at a distance of at least 0.5 meters to prevent eyestrain. Therefore, it was decided to place the displays in the \enquote{close} layout conditions (layouts EL/C and BEL/C) 0.55 meters from the user. For the distance of the layouts at a further distance, design guidelines by Apple and Meta were considered. While Meta recommends around one meter for a comfortable viewing distance, Apple uses two meters as the default distance when users open a window on their Apple Vision Pro. Based on the environment at the library, it was decided to place the displays in the \enquote{far} layout condition (layout EL/F and BEL/F) 1.7 meters from the user. At this distance, the virtual displays are projected above the desk of the person sitting across from the user, allowing us to investigate the potential effects of placing virtual displays in the personal space of bystanders. A distance of more than 1.7 meters was dismissed because this would cause the displays to collide with the person sitting at the opposite desk. The size of the virtual screens was scaled in proportion to their distance so that displays at further distances appear larger to ensure the same relative size and readability for all layout conditions. For the layouts at closer distance, the size of the screens is similar to an average PC monitor. Since the displays in the layouts at further distance are placed at around 3.1 times the distance, the scale of these displays is 3.1 times the scale of the displays in the \enquote{close} layout conditions. Based on these measures, four layouts were created in Bezi. Using the Bezi-Unity integration feature, a file link was generated to import the 3D prototype into Unity. Unity was used to implement the final interactive MR application. It allows users to enter their answers to the room planning questions and includes a menu where participants can select the different layouts. Unity offers an MR template, which was used as a starting point for the Unity project. The MR template allows the recognition of inputs from the Meta Quest controllers and implements the passthrough mode, which enables the MR experience. A \enquote{home menu} scene was created. When starting the application, the first scene a participant sees prompts the user to select one of the four layouts (see Fig. \ref{fig: prototype}-left). As a next step, buttons were added to make the prototype interactive. Check-boxes were added to the central display to allow users to enter their answers to the room planning questions, and a \enquote{continue} button was added, which users are supposed to click once they finish the task (see Fig. \ref{fig: prototype}-center). Clicking the \enquote{Continue} button causes a pop-up window to open, which prompts the user to fill out a questionnaire (see Fig. \ref{fig: prototype}-right). By clicking on the \enquote{Continue} button in the pop-up window, the user is taken back to the home menu, where they are asked to select the next layout. Lastly, the Unity project was duplicated and modified to create a separate application for the training task, which contain only one room planning question per layout.

\subsection{Participants}
In total, 22 people took part in the study. One participant was excluded from the analysis because they did not follow the instructions correctly. Additionally, two participants were excluded from further analysis because their data showed strong outliers. Out of the final sample of 19 participants, eleven were female, and eight were male. Their mean age was 27.74 (SD = 0.78, min. 23, max. 34 years). The participants got recruited from university students and by word of mouth. They were all students, with the majority being Human Factors Master students. Three participants had no prior experience with VR or MR, 4 participants had tried VR or MR once, 11 people had experienced it a few times, and 1 participant had used it many times. The sample's mean Affinity for Technology Interaction score was 4.60 (SD = 0.16, $\alpha$ = 0.86), with the minimum being 3.0 and the maximum 5.33 on a scale from 1 to 6 \cite{franke2019personal}. The main part of the study took place in an open study space in a public library. During the experiment, on average, 19 other people were present in the study area (min. 12, max. 25 bystanders). Since the order of the layouts was counterbalanced, resulting in 24 different orders, no order was experienced by more than one participant. Participants were offered test subject credits for their participation.

\subsection{Procedure}
\label{subsec: procedure}

The duration per participant varied between 50 and 70 minutes. All questionnaires were answered via a Google form on a tablet. The study was conducted in English. The first part of the study, including instructions and the training task, took place in a private room at a public library. There, the participant read a pre-printed study overview and then signed a consent form. After that, the participants answered a demographic questionnaire, a question about their previous experience with XR, and the Affinity for Technology Interaction questionnaire on a tablet. In the next step, participants were handed a pre-printed explanation of the room planning task, including the order in which they should select the layouts.
After wearing and adjusting the headset, the participants were told how to use the controller and center their view. The participants were then ready to start a training task: they went through all four layouts in the order they were given (different for each participant) and answered one room planning question per layout.

After completing the training task, the moderator and the participant went together to the library's public area. The participant was asked to take all their belongings with them. The moderator chose a desk for the participant so that a person was sitting across from the participant and as many desks as possible around the participant were occupied. After the moderator had started the application on the headset, the participant sat down at the desk. The tablet to answer the questionnaires and the sheet of paper containing the layouts' order were placed on the participant's desk. Once the participants were comfortable, they pressed the button to center the virtual screens. The participant selected the first layout, based on the order of layouts they were given, and started working on the task. 

After finishing the task, the participants used the tablet to answer the questionnaire, keeping the headset on.  They answered a post-task questionnaire based on their experience working with the current layout. It includes perceived productivity, the feeling of safety and social acceptability rated on a 7-point Likert scale:

    \vspace{1mm}
    \textbf{Perceived productivity:} it was measured with two adapted questions taken from \cite{pavanatto2021we}:\enquote{I found it easy to perform the task using this layout of virtual displays; I think that I was able to focus on my task.} A high value indicates a high level of productivity.
    
    \vspace{0.5mm}
    \textbf{Feeling of Safety:} it was measured with a single question from \cite{bajorunaite2023reality}: \enquote{I felt safe using the headset in this scenario with this layout.}
    
    \vspace{0.5mm}
    \textbf{Social Acceptability:} the Social Acceptability Questionnaire (SA) was taken from a previous study by Eghbali et al. \cite{eghbali2019social}, who created an adapted version of a social acceptability questionnaire proposed in \cite{profita2016effect}. The questionnaire by Eghbali et al. consists of 11 statements sorted into seven thematic groups. For this study, some formulations were slightly modified, and three items were removed that were not applicable in the context of the study, resulting in the following eight questions:
    \begin{itemize}
    \vspace{-1mm}
        \item \textit{Public Use}: \enquote{It felt appropriate to use the Mixed Reality headset in this public place; It felt rude to use the Mixed Reality headset in this public place; It felt uncomfortable being watched by others while using the Mixed Reality headset in this public setting}
    \vspace{-2mm}
    \item \textit{User}: \enquote{I think the Mixed Reality headset makes me look cool}
    \vspace{-5mm}
    \item  \textit{Interaction}: \enquote{It felt awkward doing head movements while using Mixed Reality in public}
    \vspace{-2mm}
    \item \textit{Isolation}: \enquote{I did not like the fact that I was isolated from the rest of the people in the public space} 
     \vspace{-2mm}
    
    \item \textit{Privacy}: \enquote{I was concerned about spectators recording me while using Mixed Reality in public}
     \vspace{-2mm}
    \item \textit{Safety}: \enquote{I was concerned about bumping into objects or people while using Mixed Reality in public}
    
    \end{itemize}
    

While the participants were working on their tasks, the moderator sat down at another desk in the study area so that they could observe the participant from a distance. The moderator noted how many people are in the study area at the time and if something abnormal happened or the person sitting across from the participant left. The moderator used the casting functionality in-between to quickly check whether the participants were selecting the layouts in the correct order but did not observe the participants during the task. The same procedure was repeated for the other three layouts. Finally, the participant was asked to fill out a final questionnaire. The final questionnaire included a mix of closed- and open-ended questions. First, participants were asked to rank the four layouts in order of preference and explain their ranking. To further explore the factors that impact layout preferences, participants were asked to rate the following three questions on a 5-point Likert scale: 1. \enquote{Is it important to you that the virtual displays don't block your view of the people around you?} 2. \enquote{Is it important to you that the virtual displays don't collide with physical objects in the real word (for example, the lamp or table)?} 3. \enquote{Is it important to you that the virtual displays are not positioned within the personal space of other people?} Lastly, participants are asked if they would use MR in public again, what they liked and disliked, and how using MR in public could be improved.

\section{Results}

In this section, the results from the repeated measures ANOVA and contrast analysis are presented. Despite the test of normality showing non-normal distribution for several dependent variables, it was chosen to perform a repeated measures ANOVA because it is quite robust to violations of normality \cite{vasey1987continuing, berkovits2000bootstrap,blanca2023non}. The sphericity assumption was met for all dependent variables except one; in this case, the Greenhouse-Geisser correction was applied. We used two one-sided t-tests (TOST) for testing the equivalence between conditions. The equivalence lower and upper bounds were calculated using Cohen’s $d_z = \frac{t}{\sqrt{n}}$ with the critical t value for n = 19, $\alpha$ = 0.05 for two-tails (cf. \cite{lakens2018equivalence}). The value of $d_z$ for our analysis is $\pm 0.48$ with $ -0.48 $ as the lower bound and $ +0.48 $ as the upper bound. The significant equivalence tests are reported in Table \ref{tab:TOST}.

 \vspace{-1mm}
\begin{table}[htbp]
  \centering
   \caption{One-way repeated-measures ANOVA results}{
   \vspace{-0.3cm}
\begin{tabular}{lcclrc}
\toprule
        Effect & \( df_{n} \) & \( df_{d} \) & \( \hphantom{F}F \) & \( p\hphantom{p.} \) & \(\eta_{G}^{2} \) \\
\midrule
        
        Easiness   
        & 3	& 54 & 1.843 & .150 & .093  \\
        Focus   
        & 3	& 54 & 5.377 & .003	& .230  \\
        Safety
        & 3	& 54 & 2.879 & .044	& .138 \\
        SA\_Public 
        & 3	& 54 & 1.418 & .247 & .073  \\
        SA\_User   
        & 3	& 54 & 1.385 & .257 & .071  \\
        SA\_Interaction   
        & 3	& 54 & .685 & .565 & .037  \\
        SA\_Isolation      
        & 3	& 54 & .186 & .906 & .010  \\
        SA\_Privacy          
        & 3	& 54 & .431	& .732	& .023  \\
        SA\_Safety 
        & 1.899	& 34.177	& .703 & .495 & .038  \\

\bottomrule
    \end{tabular}%
    }
  \label{tab:ANOVAresults}%
  \vspace{-0.3cm}
\end{table}%

\begin{table}[htbp]
  \centering
   \caption{ Contrasts test results}{
   \vspace{-0.3cm}
\begin{tabular}{lcclrc}
\toprule
        Effect & \( df_{n} \) & \( df_{d} \) & \( \hphantom{F}F \) & \( p\hphantom{p.} \) & \(\eta_{G}^{2} \) \\
\midrule
        
        Easiness   
        & 1 & 18 & 3.109 & .095	& .147  \\
        Focus   
        & 1 & 18 & 12.565 &	.002 & .411  \\
        Safety
        & 1	& 18 & 3.473 & .079	& .162 \\
        SA\_Public 
        & 1	& 18 & 4.357 & .051 & .195  \\
        SA\_User   
        & 1	& 18 & .445	& .513 & .024  \\
        SA\_Interaction   
        & 1	& 18 & 1.306 & .268	& .068  \\
        SA\_Isolation      
        & 1 & 18 & .091 & .767 & .005 \\
        SA\_Privacy          
        & 1	& 18 & .151	& .702	& .008  \\
        SA\_Safety 
        & 1	& 18 & .035 & .853 & .002  \\
\bottomrule
    \end{tabular}%
    }
  \label{tab:contrastsresults}%
  \vspace{-0.5cm}
\end{table}%

\begin{table*}[htbp]
  \centering
   \caption{Statistically significant TOST results for testing equivalence. For each layout pair the t-value, and the larger between lower and upper bound p-values are reported for the social acceptability effects.}
   \vspace{-0.2cm}
\begin{tabular}{lllllll}
\toprule
        Effect / Layouts & EL/C,EL/F & EL/C,BEL/C & EL/C,BEL/F	& EL/F,BEL/C & EL/F,BEL/F & BEL/C,BEL/F \\

\midrule
        
        SA\_User   
        & t(18) = -1.71 & - & - & t(18) = -.697 & t(18) = 0 & t(18) = .697  \\
        & p = .021 & - & - & p = .012 & p = .006 & p = .012 \\
        \hline

        SA\_Privacy          
        & - & t(18) = .271 & - & - & - & -  \\
        & - & p = .021 & - & - & - \\
        \hline
        SA\_Safety 
         & t(18) = -.369 & t(18) = .524 & t(18) = -.567 & t(18) = 1.14 & t(18) = -.438 & t(18) = -2.19  \\
         & p = .004 & p = .039 & p = .029 & p = .016 & p = 0.001 & p = .006 \\
        
\bottomrule
    \end{tabular}%
  \label{tab:TOST}%
  \vspace{-0.5cm}
\end{table*}%

\subsection{Productivity}

Regarding the easiness of performing the task, layout EL/C was rated the highest (M = 5.21, SD = 1.08), followed by EL/F (M = 5.16, SD = 1.46), BEL/C (M = 4.95, SD = 1.08) and BEL/F (M = 4.42, SD = 1.50). A repeated measures ANOVA showed no statistically significant difference in easiness between the different layout conditions (see Table \ref{tab:ANOVAresults}). A contrast analysis revealed that there is also no significant difference between layouts on eye level (EL/C and EL/F) and layouts below eye level (BEL/C and BEL/F) (see Table \ref{tab:contrastsresults}). No equivalence was found for each pair of layouts. 

For the ability to focus on the task, layout EL/C was rated the highest (M = 5.74, SD = 1.05), followed by EL/F (M = 5.37, SD = 1.46), BEL/C (M = 4.89, SD = 1.29) and BEL/F (M = 4.42, SD = 1.58). A significant difference in the ability to focus was found between the different layout conditions (see Table \ref{tab:ANOVAresults}). The ability to focus was significantly higher (see Table \ref{tab:contrastsresults}), for layouts on eye level (layouts EL/C and EL/F) than for layouts below eye level (layouts BEL/C and BEL/F), with a mean difference of 0.90 (SE = 0.25). No equivalence was found for each pair of layouts.

\subsection{Safety}

Feeling of safety was the highest for layout EL/C (M = 5.53, SD = 1.12), followed by BEL/F (M = 5.47, SD = 1.07), EL/F (M = 5.37, SD = 1.02) and BEL/C (M = 4.74, SD = 1.28). A significant effect of layout on participants’ rating of safety was found (see Table \ref{tab:ANOVAresults}). However, in the post-hoc contrasts test the difference between layouts BEL and layouts on EL was not significant (see Table \ref{tab:contrastsresults}). No equivalence was found for each pair of layouts.

\subsection{Social Acceptability}

For reliability analysis, Cronbach’s alpha was calculated to assess the internal consistency of the subscale \enquote{statements about public use}, which consists of three questions. Cronbach’s alpha was calculated separately for all four layout conditions in which this subscale was used. The internal consistency of the questionnaire proved to be acceptable for all conditions (for layout condition EL/C: $\alpha$ = 0.80, for layout condition EL/F: $\alpha$ = 0.72, for layout condition BEL/C: $\alpha$ = 0.69, for layout condition BEL/F: $\alpha$ = 0.60). For the statements about public use, the social acceptability was highest for layouts EL/C (M = 4.68, SD = 1.26) and EL/F (M = 4.68, SD = 1.64), followed by BEL/C (M = 4.28, SD = 1.35) and BEL/F (M = 4.19, SD = 1.31). 
Participants thought the headset made them look the coolest when working with layout EL/C (M = 2.53, SD = 1.54), followed by EL/F (M = 2.32, SD = 1.34), BEL/F (M = 2.32, SD = 1.25) and BEL/C (M = 2.21, SD = 1.23).  Regarding the statement about awkwardness of head movements, the participants felt least awkward when using layout EL/C (M = 4.26, SD = 1.45) and BEL/F (M = 4.26, SD = 2.00), followed by EL/F (M = 3.84, SD = 1.64) and BEL/C (M = 2.21, SD = 1.75). For the aspect of isolation, the social acceptability was highest for layout EL/F (M = 4.89, SD = 1.66), followed by BEL/C (M = 4.84, SD = 1.74), EL/C (M = 4.74, SD = 1.82) and BEL/F (M = 4.68, SD = 1.19). Results show that the participants were least concerned about being recorded by spectators when using layout BEL/C (M = 5.47, SD = 1.61), followed by EL/C (M = 5.42, SD = 1.61), EL/F (M = 5.37, SD = 1.26) and BEL/F (M = 5.16, SD = 2.01). Regarding the concern of bumping into objects or people, the participants were least concerned when using layout BEL/C (M = 6.53, SD = 0.61), followed by EL/C (M = 6.42, SD = 0.84), EL/F (M = 6.37, SD = 0.68) and BEL/F (M = 6.32, SD = 0.75). No significant difference between the different layouts was found for any aspect of social acceptability (see Table \ref{tab:ANOVAresults}). The difference between layouts on eye level (layouts EL/C and EL/F) and below eye level (layouts BEL/C and BEL/F) was also not statistically significant for all social aspects (see Table \ref{tab:contrastsresults}). TOST indicated equivalence between pair of layouts for the User, Privacy and Safety dimensions of social acceptability (see Table \ref{tab:TOST}).

\subsection{Layout Preferences}

In terms of layout preferences, layout EL/F (eye level-far) was most preferred for working at the library, with a mean preference score of 7.16 (SD = 3.24), on a scale from 1 to 10, where a preference score of 10 would mean that all participants rated it as most preferred. The second most preferred layout was layout EL/C (eye level-close) (M = 5.89, SD = 3.64), followed by layout BEL/F (below eye level-far) (M = 4.95, SD = 3.01). On average, layout BEL/C (below eye level-close) was least preferred (M = 4.00, SD = 3.00). The participants who preferred eye level layouts reasoned that they preferred not to see the people around them (P1, P2, P3, P4, P5, P8, P11, P12, P16). Participants explained that not seeing others is less distracting (P1, P8, P12) and more comfortable (P2, P5). Additionally, one participant mentioned that they found it awkward to be able to look the person sitting across in the eyes (P3). Two participants stated that they preferred the layouts on eye level because it allowed for a more comfortable head position (P2, P18), with one participant further explaining that the headset's weight pushes downwards (P2). Users whose favorite layout was layout EL/C reasoned that it feels most similar to working with a setup of physical monitors (P16) and because it allows for the right amount of awareness of the surroundings (P18). The participants who favored the layouts below eye level reasoned that they like to see their surroundings (P6, P7, P10), with one participant explicitly mentioning that it makes them feel safer in a public place (P6). One participant said that they preferred the layouts below eye level because the background behind the virtual screens is steadier since there are no moving objects or people visible in the gaps between the screens (P13). Those who preferred the layouts at a further distance argued that it feels more comfortable than the closer screens (P2, P3, P6, P17), with one participant further explaining that it feels less restricting (P2). However, users who preferred the layouts at shorter distances also reasoned that it felt more comfortable for them (P1, P5). While one person mentioned that it is easier to view all the information at once with the layouts that have the screens further away (P2), another person felt the layouts at a shorter distance gave them a better overview (P5). Another factor influencing participants' preference rankings was how much they had to move their heads when working with a specific layout. Participants preferred when they had to move their heads less. However, two participants stated fewer head movements for preferring layouts at further distances (P3, P14), while two other users mentioned it as a reason for preferring layouts at closer distances (P4, P12). One participant reported that moving the head feels weird and makes concentrating harder (P12).
Regarding the question of how important it is that the virtual displays do not block the view of other people, 11 participants (58\%) found it rather unimportant or not important at all. In comparison, 6 participants (32\%) found it rather important or very important. Regarding the question about the importance of virtual displays not colliding with physical objects, ten participants (53\%) found it rather important or very important, while 8 participants (42\%) found it rather unimportant or very important. When asked how important it is that the virtual displays are not positioned within other people's personal space, 9 (47\%) participants responded that they found it rather important or very important, while 8 (42\%) participants found it rather unimportant or not important at all. 

\subsection{Using MR in public}

When participants were asked what they liked about the MR experience, six people reported that it was interesting since it was a novel experience (P1, P2, P3, P13, P15). Participants especially liked having multiple screens (P8, P9, P12, P14, P16) and found it comfortable and efficient to work with (P8, P12, P14). One participant mentioned that they appreciated the variety of possible screen sizes and placements (P7). Some participants liked that MR allowed them to be still aware of their surroundings (P4, P10, P18), while other participants liked feeling isolated from other people (P5, P6). One participant pointed out that using MR allowed them to concentrate better while still retaining some situational awareness (P18). Additionally, it was mentioned that the library's location felt safe (P6) and that using MR in public was not as uncomfortable as expected (P2). Lastly, one participant reported that they appreciated the display quality (P6), and another liked that MR caused less motion sickness than VR (P4). Regarding the things participants did not like about the MR experience, discomfort was a major factor. Participants disliked the display quality (P2, P7, P9, P10, P13, P15, P18) and said that it made them feel dizzy (P2, P10, P15) or made their eyes hurt slightly (P7). Furthermore, some participants mentioned that they found it uncomfortable to wear the headset (P2, P3, P14), for example, because of the weight and pressure (P2, P14). One participant reported that they found it uncomfortable to make many large head movements to view the screen (P6). Another aspect that participants did not like is the uncomfortable or awkward feeling of using MR in public (P1, P3, P5, P8, P16, P17). Some felt uncomfortable because they were the only person using MR (P8) and felt watched (P3). Some participants reported concerns about social acceptability from the bystander's point of view (P1, P4, P18). For example, participants were worried about invading other people's personal space (P4) and making others feel uncomfortable (P1, P4) because others might not be aware that the user can see them through the headset (P1) or because they could feel like they are being watched or filmed (P4). One participant mentioned that they did not like pointing at bystanders when using the controller in layout condition EL/C (on eye level and close) (P18). Furthermore, two participants mentioned that they disliked the virtual display colliding with physical objects in the environment (P6, P8). Lastly, another participant said they found it difficult to visually search for relevant information when using the layouts with the bigger screens at a further distance (Layouts EL/F and BEL/F) (P12). When asked how the experience of using MR in public could be improved, multiple participants suggested using it in places where other people also use XR (P3, P5, P15, P17) and using it somewhere where no person is sitting opposite from them (P3, P16). Additionally, it was mentioned that a less noticeable headset design could improve the experience (P5). Multiple participants suggested improving the headset's resolution (P2, P7, P8, P13, P18), and two participants proposed improving the headset's comfort (P7, P14), for example, by making it lighter (P7). Furthermore, participants mentioned that more subtle navigation mechanisms, which require less hand (P4, P18) and head movement (P5), could improve the user experience. Some other suggestions include adding an option to easily turn the virtual screens transparent (P6), including other senses, for example, by adding haptic feedback (P19) and having the option to fade out other people so that only the objects in the physical environment are visible (P11). Of the 19 participants, six reported they would use MR in public again, two said they would not, and 11 responded with \enquote{maybe.}

\section{Discussion}
\subsection{Effects of Layouts on Feeling of Safety, Social Acceptability and Productivity}

While the first hypothesis (H1) predicted that users feel safer with layouts BEL than layouts on EL, the results do not support this hypothesis, since no statistically significant difference between the two height levels was found. In fact, participants felt the safest with layout EL/C, which positions displays on eye level so that the view of bystanders is blocked, and felt least safe with layout BEL/C, for which users can see bystanders above the virtual displays. This is contrary to prior research, which suggests that users feel safer when they can see more of their surroundings \cite{bajorunaite2023reality, biener2024}. It must be noted that those studies did not compare different MR layouts, but rather compared MR and VR and conducted the studies in different types of environments. It is to be expected, that the setting in which XR is used influences user’s ratings of feeling of safety. Overall, the mean safety values show that participants felt rather safe for all layouts. This implies that the library is a space where people do not expect to encounter dangerous situations. It can be assumed that, not only the level of safety, but also the kind of safety concerns that people have depend on their environment. For example, in a public transport setting people are likely to be more concerned about theft compared to at a library. If in previous studies the increased sense of safety from having more visibility of the surroundings were influenced by concerns that are not as relevant in a library setting, such as theft, this might explain why having more visibility of the surroundings in the library did not make people feel safer. However, a more differentiated measure of safety would be needed to test this speculation. A possible explanation for why users could even feel less safe at the library when using layouts which provide visibility of bystanders, is that they feel more exposed to the gaze of other people. It is conceivable that participants rated the feeling of safety for layouts BEL/C and BEL/F (below eye level) lower compared to layout EL/C (on eye level) because the layouts below eye level made them more aware of the presence of other people and therefore could make them feel more watched and more aware of possible threats. Nonetheless, the qualitative data suggests that feeling of safety is subject to individual differences, since one participant explicitly mentioned that being able to see their surroundings made them feel safer, even though this was not a general trend in the quantitative data.

Regarding the second hypothesis (H2), which predicted that EL layouts make users feel more productive than BEL, the results show that this can only be confirmed with regards to the aspect of being able to focus. The ability to focus was significantly higher for EL layouts than for BEL ones with a large effect size. This finding is backed up by qualitative data, since multiple participants explained that seeing other people is distracting. This result is in line with previous research, which suggests that people feel less distracted when they can see less of their surroundings \cite{bajorunaite2023reality, lee2022partitioning}. Regarding the second aspect of productivity, the perceived easiness of performing the task, the EL layouts were rated higher than the BEL ones, but the difference was not significant. Even though the task questions for each layout were designed to have the same difficulty level and it was tried to avoid learning effects by introducing new task documents for each layout, it is not possible to ensure that no factors other than the layout differences impact the participant’s responses.

With respect to social acceptability, it was found that the differences in the layouts did not have an effect on any of the measured aspects of social acceptability. Therefore, H3, which predicted that BEL layouts would be perceived as more socially acceptable than EL layouts, is rejected. Thus, findings from previous research, which suggest that users perceive using an XR headset as more socially acceptable when they can see more of their physical surroundings \cite{bajorunaite2023reality, medeiros2022shielding}, could not be replicated. Both of those studies used VR to simulate the experience of MR in a public transport setting. Reasons for the varying results could be differences between simulations and actual public spaces or the different kinds of public settings. On the one hand, it is possible that, when simulating public settings with virtual avatars, participants do not grasp the social aspects of the experience to the same extent as they would in real public settings with actual bystanders. Even though no significant difference was found, the results show that, contrary to the hypothesis, the mean social acceptability ratings for statements about public use were higher for layouts on eye level than for layouts below eye level. A possible explanation is that, when using layouts that block the view of their surroundings, users are less aware of bystanders' presence and may be less mindful of social acceptability concerns. Such effects might be difficult to measure in simulations of public settings. Even though this study did not find any significant differences in social acceptability between the different virtual layouts, the results might differ if the same study were conducted in other public settings, where other social norms apply. For example, layouts that block the view of user’s surroundings could be considered less socially acceptable in public settings where users are more likely to interact with bystanders. It remains unclear, whether the deviations from previous research results stem from simulation effects, differences in the type of public settings, a combination of both or other differences in the study design.

\subsection{Layout Preferences}

This study addresses the question of which layouts are most preferred by users in the library setting (RQ1). The results show that, on average, the layout with displays on eye level and further away (layout EL/F) was most preferred. In general, both layouts on eye level were preferred over the layouts below eye level. In addition, it should be noted that the standard deviation of the mean preference scores is quite high for all layouts, indicating that layout preferences vary substantially among individuals. This is in line with previous research, which observed that layout preferences are not universal \cite{cheng2021semanticadapt,medeiros2022shielding}. Reasons for individual differences could for example be different character traits like the tendency to feel anxious or uncomfortable in social situations. Lastly, this study contributes to understanding the reasons behind layout preferences (RQ2). Most reasonings were related to the visibility of bystanders and the associated distractibility or aspects of social, physical, and visual comfort. Lower distractibility and higher social comfort were mentioned as explanations for preferring layouts on eye level. Regarding the physical and visual comfort of the virtual layouts, participants were not in agreement about which layouts are more comfortable. It is notable that safety was rarely mentioned in the explanations for layout preferences, indicating that safety concerns do not play a large role at the library, possibly because it is a place where people tend to feel safe. The observed preference for layouts on eye level can be explained by the finding that people find it rather unimportant whether virtual displays block the view of other bystanders, and many users stating that they actually prefer to not see the people around them. This finding may appear contradictory to prior research which indicates a preference for virtual layouts that do not obstruct the view of bystanders \cite{medeiros2022shielding}. However, given that the mentioned study was focused on public transport settings, this discrepancy aligns with previous findings suggesting that preferences for layouts are influenced by the environment in which they are used \cite{medeiros2022shielding,ng2021passenger}. Both qualitative and quantitative results of this study suggest that in a library environment, which people usually visit with the intention of being productive, users favor layouts on eye level because they are less distracting. Therefore, this study underscores the importance of considering the type of public setting when designing MR layouts, as preferences may vary widely across different environments. Furthermore, it is possible that participants in previous research \cite{medeiros2022shielding} underestimated how distracting and uncomfortable it is to see bystanders in MR, since the study simulated public settings with avatars in VR and participants did not interact with the virtual layouts for a prolonged period. This would imply that layouts which obstruct the view of bystanders may generally not be as unfavorable as previously suggested. Nevertheless, this study also suggests that preferences regarding the visibility of bystanders are not fully consistent across individuals.

\subsection{Implications for the design of MR interfaces} 
The fact that the majority of participants stated that they would consider using MR in public again emphasizes that it is worthwhile to explore how MR interfaces could be optimized for the use in public spaces. Based on the aforementioned individual differences, it can be concluded that it is not feasible to find a MR layout that satisfies everyone’s preferences. Therefore, designers of MR interfaces should give users some freedom for customizing layouts according to their personal preferences. Users should be able to adjust the height and distance at which displays are placed, like it is also common in current headsets by Meta and Apple. Nevertheless, it is usually necessary to define a default layout. These defaults should be chosen so that the user’s need for adjustments is minimized. As research indicates that layout preferences depend on the settings in which they are used, the ideal choice for a default layout also varies based on the environment. In a library environment, where the goal is to be productive, it could be advisable to use a default layout on eye level, since blocking the view of bystanders helps with the ability to focus. More specifically, this research suggests that a layout arranging displays on eye level at a further distance would be the preferred default layout at a library workspace for most people. This default layout may also work well in other public environments where the goal is to be productive, users feel safe, similar social norms apply and the seating arrangements are similar. This study focused on a coworking setting in a public library. While comparable results might be expected in similar environments, it is strongly advisable to study further settings in future studies. In other settings, like on a subway, where other social norms apply and more situational awareness is needed, the ideal default layouts might be one that arranges displays below eye level. 
Still, our investigations can be seen as a first step towards a better understanding of using XR-technologies in public spaces. Additionally, all default layouts should avoid collisions between the virtual displays and physical objects. These recommendations for default layouts could be used to inform the development of algorithmic approaches for automated layout adaption (similar to \cite{cheng2023interactionadapt}). 
\vspace{-1mm}
\section{Limitations}

Reflecting on the study methodology, it is worth noting that the item assessing the ease of performing the task may not reliably measure effects of layouts on perceived productivity, as it is difficult to ensure that no factors other than the layout differences influence participants’ responses.  Also, while the high ecological validity increases the external validity, the generalizability of the results to other populations is limited due to the participant pool. In fact, most of the participants recruited were university students, they are a homogenous group in terms of age, education, level of technological affinity and prior experience with XR. Finally, the results might not apply to other public settings because the feelings of safety, social acceptability, and productivity may vary depending on the setting.

\section{Conclusion and Future Work}

This study compares four different layouts (EL/C, EL/F, BEL/C, BEL/F) of virtual displays regarding feelings of safety, perceived productivity, and social acceptability when working with MR in public. The participants (N=19) were exposed to the different layouts and engaged in a room-planning task while wearing an MR headset in a public library. The results from a repeated measures ANOVA rejected the hypothesis (H1) that users feel safer with BEL layouts and the hypothesis (H3) that BEL layouts are more socially acceptable. Besides, they partially confirmed the hypothesis (H2) that EL layouts are perceived as more productive than BEL ones by showing statistically significant results in the focus dimension of productivity. Furthermore, the results indicate that layout EL/F was the most preferred and that, in general, the visibility of bystanders influences the layout preferences. 

In this research, deviations from previous research findings about MR in other settings were identified, but it remains unclear to what extend these deviations are attributable to the different environments or disparities in the study design, particularly the usage of VR to simulate public settings in previous studies. By examining different public settings, one could test the hypothesis, that layouts which obstruct user’s surroundings would be perceived as less socially acceptable in settings where users are likely to interact with bystanders. 
Some participants expressed concerns about the social acceptability from the bystander’s perspective, so it would be interesting to include bystanders in future surveys about MR layouts. Also, specific hypotheses on objective metrics (e.g., eye-gaze points or fixation) could add valuable information to the researched topic. Lastly, given that several participants mentioned feeling uncomfortable when using MR in public, it would be interesting to investigate whether these feelings diminish with regular use. It would be valuable to examine such long-term effects on social acceptability, since this could influence layout preferences.

\newpage

\bibliographystyle{abbrv-doi}

\balance
\bibliography{main}
\end{document}